\documentclass[11pt]{article}
\usepackage[margin=1in]{geometry}
\usepackage{amsmath,amssymb,amsthm,mathtools,bm,booktabs,graphicx}
\usepackage[colorlinks=true,linkcolor=blue,citecolor=blue,urlcolor=blue]{hyperref}

\newtheorem{theorem}{Theorem}
\newtheorem{proposition}[theorem]{Proposition}
\newtheorem{lemma}[theorem]{Lemma}
\newtheorem{corollary}[theorem]{Corollary}
\newtheorem{conjecture}[theorem]{Conjecture}
\theoremstyle{remark}
\newtheorem{remark}[theorem]{Remark}

\newcommand{\dg}{^\dagger}
\newcommand{\ad}{a\dg}
\newcommand{\Hc}{H_{\rm c}}
\renewcommand{\SS}[1]{\mathfrak S_{#1}}
\newcommand{\ket}[1]{|#1\rangle}
\newcommand{\bra}[1]{\langle#1|}
\newcommand{\dom}{\operatorname{dom}}
\newcommand{\ran}{\operatorname{ran}}
\newcommand{\tr}{\operatorname{tr}}

\begin{document}

\title{Structural Spectral Determinants for Coupled Quantum Systems}
\author{Ghasem Asadi Cordshooli\\[2pt]
\small Department of Physics, YI. C., Islamic Azad University, Tehran, Iran\\
\small \texttt{ghascor@iau.ac.ir}}

\date{}
\maketitle

\begin{abstract}
Quantum Hamiltonians decomposable into finitely many coupled channels associated with internal states have largely been studied on a model-by-model basis. In the present work, a structural spectral criterion is established by retaining one channel and treating all remaining channels as a single coupled block through a Frobenius--Schur reduction. Under suitable conditions on the channel operators and their couplings, the eigenvalues of the full Hamiltonian outside the spectra of the retained channel and the coupled block are exactly the zeros of a regularized Fredholm determinant. The order of each zero equals the multiplicity of the corresponding eigenvalue. For channels unitarily equivalent to harmonic oscillators, the required regularization order is determined by the number of oscillator modes and the growth of the interchannel couplings. Matching lower bounds establish the optimality of this order for bounded displacement and squeezing couplings in the oscillator normal form, as well as for linear single-mode couplings. The framework encompasses the Friedrichs--Lee, Rabi, Jaynes--Cummings, Rabi--Stark, two-photon, and two-mode models, a three-level emitter, the Dicke model, and arrays of inhomogeneous qubits. In the Rabi--Stark model, the operator conditions select the bare-qubit decomposition, while in the quantum Rabi model the determinant factorizes into parity-resolved factors.
\end{abstract}

\section{Introduction}
\label{sec:intro}

Many quantum Hamiltonians decompose into finitely many coupled channels, each associated with an internal state of the system, and their spectra have largely been obtained model by model. This is particularly evident in light--matter and electron--phonon systems, where the spectral problem may involve several interacting internal states, oscillator modes, or discrete configurations. Although the underlying Hamiltonians share a block structure, their exact spectral descriptions have generally been developed through constructions adapted to the algebraic properties of each model. The question addressed here is whether this common operator structure provides a systematic route to spectral characterization.

The search for exact spectra has a long history in quantum light--matter models. A major development was Braak's solution of the quantum Rabi model (QRM), in which the discrete $\mathbb{Z}_2$ symmetry is used to construct an analytic $G$-function in Bargmann space whose zeros determine the regular spectrum~\cite{Braak2011}. Subsequent approaches based on extended coherent states, squeezed states, and Bogoliubov-operator expansions provided alternative descriptions of the Rabi spectrum and extensions to nonlinear and multimode variants~\cite{Chen2012,Moolekamp2012,Braak2013CF}. Related methods have been developed for the Dicke model, two-qubit Rabi systems, and other multilevel and multimode extensions~\cite{BraakDicke2013,TwoQubitRabi,BatchelorZhou2015,Duan2015TwoQubit,xie2019}. The two-photon Rabi model and its generalizations have also received particular attention because of their spectral-collapse regimes, where infinitely many discrete levels accumulate at a critical energy~\cite{Travenec2012,MaciejewskiComment2015,Felicetti2018,Duan2022,Braak2023TwoPhoton,Braak2024Correction}. More recent constructions use $su(1,1)$ representations and recurrence relations to treat several nonlinear Rabi models within related analytic schemes~\cite{Duan2022}. Spectral determinants and zeta functions associated with the Rabi family have further been investigated as mathematical objects in their own right~\cite{ReyesBustosWakayama2021,KimotoRBW2021,HiroshimaShirai2025}.

Exact spectral constructions extend beyond cavity quantum electrodynamics. In electron--phonon systems, coherent-state expansions and diagonalization procedures have yielded exact spectra for symmetric and asymmetric two-site Holstein models~\cite{Holstein1959,han2002,wang2006b}. In the two-site Holstein model the electronic sites form two channels whose diagonal blocks are displaced phonon oscillators coupled by the electronic hopping, the same channel form as the quantum Rabi model in its displaced-oscillator basis. These developments, together with the theoretical and experimental progress in the ultrastrong light--matter coupling regime~\cite{FornDiaz2019,FriskKockum2019,Blais2021}, illustrate the range of physical settings in which nonperturbative spectral information is required. At the same time, the analytic functions and recursive constructions used in these studies are usually tied to particular symmetries, representations, or coupling patterns. A spectral criterion formulated directly in terms of the channel operators would provide a complementary description, independent of a model-specific algebraic solution.

Such a criterion requires more than the formal elimination of one component of a wave function. For a two-channel Hamiltonian, the elimination is expressed through the resolvent of the other channel. In a multichannel system, however, the eliminated channels generally remain coupled to one another. They must therefore be treated as a single complementary block, whose spectrum determines where the reduction is admissible. This distinction is important for multilevel emitters, the Dicke model, and arrays of interacting qubits, where reducing the problem to independent channel resolvents would discard part of the internal coupling structure. The appropriate reduction is the Frobenius--Schur complement, closely related to the Feshbach projection method~\cite{Feshbach1958}. It retains the internal dynamics of the complementary block while producing an effective operator on the retained channel.

The operator-theoretic ingredients of this construction are established. The Birman--Schwinger principle converts spectral questions into eigenvalue problems for compact operator families~\cite{Birman1961,Schwinger1961,Kato1966,KonnoKuroda1966}. Feshbach maps, Schur complements, and block operator matrices provide general methods for reducing coupled operator problems~\cite{BachEtAl2003,Zhang2005,Tretter2008,GriesemerHasler2008}. Their combination with the Birman--Schwinger principle has also been used for spin--boson operator matrices in truncated Fock space~\cite{IbrogimovTretter2018}. When the resulting operator belongs to a Schatten ideal, a regularized Fredholm determinant converts the operator-valued spectral condition into a scalar analytic function~\cite{Simon2005,GLM2005,Rasulov2014,Rasulov2023}. Generalized Birman--Schwinger principles and Weinstein--Aronszajn-type formulas further relate determinant zeros to eigenvalues and their algebraic multiplicities~\cite{GLM2005,HansmannKrejcirik2020,BehrndtGesztesy2020}, and an eigenvalue multiplicity formula is available for the Schur complement of a $3\times3$ block operator matrix~\cite{MuminovRasulov2015}.

The present work develops a channel-based spectral criterion from these ingredients and identifies the additional structure supplied by the reduction. A single channel is retained, while all remaining channels are eliminated together as a coupled block. On the common resolvent set of these two blocks, the spectral problem is expressed through a regularized Fredholm determinant acting on the retained channel space. Under the stated operator assumptions, its zeros coincide with the eigenvalues of the full Hamiltonian, and the order of each zero equals the multiplicity of the corresponding eigenvalue. The multiplicity result follows from the strict negativity of the energy derivative of the Schur complement. Thus the reduction provides both a scalar spectral condition and a precise relation between the analytic multiplicity of its zeros and the spectral multiplicity of the Hamiltonian.

A second question concerns the regularization required to define this determinant. The reduced kernel contains two off-diagonal coupling factors together with the resolvents associated with the retained and complementary channels. This product structure can improve its Schatten-class membership relative to the full-space Birman--Schwinger kernel, effectively halving the relevant Schatten index. For channels unitarily equivalent to harmonic oscillators, displacement and squeezing transformations yield normal forms in which the channel resolvents are diagonal in a Fock basis. Their decay, together with the growth of the interchannel couplings, determines the Schatten class of the reduced kernel and hence the required regularization order. Matching lower bounds establish the optimality of this order for bounded displacement and squeezing couplings in the oscillator normal form and for linear single-mode couplings. The result therefore specifies a sufficient regularization and, in these classes, the smallest one allowed by the operator structure.

The criterion is applied to a range of coupled-channel Hamiltonians from light--matter, multilevel, and discrete--continuum physics. The Friedrichs--Lee model illustrates the reduction with a continuum channel and a finite-rank kernel. In the quantum Rabi model, the reduced kernel is trace class and the determinant factorizes into parity-resolved factors. The Jaynes--Cummings, Rabi--Stark, two-photon, and two-mode models exhibit how different coupling structures and oscillator normal forms affect the Schatten properties of the reduced kernel. In the Rabi--Stark model, the operator conditions select the decomposition into bare qubit states. In the two-photon model, the oscillator normal form remains available only up to the spectral-collapse coupling. The three-level emitter makes explicit the role of the spectrum of the coupled complementary block, while the Dicke model and arrays of inhomogeneous qubits extend the construction to larger channel systems.

The paper is organized as follows. Section~\ref{sec:general} introduces the channel setting and establishes the determinant criterion. Section~\ref{sec:mult} proves the multiplicity formula. Sections~\ref{sec:osc} and~\ref{sec:order} construct the oscillator kernels and determine their regularization orders. Section~\ref{sec:analytic} studies analyticity, singularities, and truncation. Section~\ref{sec:models} presents the model applications, including the computational cost of the determinant for the quantum Rabi model in Sec.~\ref{sec:rabi}.

\section{Coupled channels and the Schur reduction}
\label{sec:general}

This section states and proves the determinant criterion.

\subsection{Setting}
The Hilbert space and the Hamiltonian are
\begin{equation}
\mathcal H=\bigoplus_{i=1}^{r}\mathcal H_i,\qquad
H=H_0+V,\qquad
H_0=\operatorname{diag}(H_1,\dots,H_r),
\label{eq:setting}
\end{equation}
with $r<\infty$, channel Hamiltonians $H_i$ on $\mathcal H_i$, and an off-diagonal coupling $V$ with blocks $V_{ij}:\mathcal H_j\to\mathcal H_i$. Channel $1$ is retained. With
\begin{equation}
W=\begin{pmatrix}V_{12}&\cdots&V_{1r}\end{pmatrix},\qquad
\Hc=\begin{pmatrix}H_2&\cdots&V_{2r}\\ \vdots&\ddots&\vdots\\ V_{r2}&\cdots&H_r\end{pmatrix},
\label{eq:blocks}
\end{equation}
the operator $H-E$ takes the form
\begin{equation}
H-E=\begin{pmatrix}H_1-E&W\\ W\dg&\Hc-E\end{pmatrix}.
\label{eq:twoblock}
\end{equation}
The complementary block $\Hc$ carries the mutual couplings of channels $2,\dots,r$. With $R_1(E)=(H_1-E)^{-1}$ we define
\begin{equation}
\Sigma(E)=W(\Hc-E)^{-1}W\dg,\qquad
K(E)=R_1(E)\,\Sigma(E),
\label{eq:Kdef}
\end{equation}
on the set
\begin{equation}
\mathcal D=\mathbb C\setminus\bigl[\sigma(H_1)\cup\sigma(\Hc)\bigr],\qquad
\mathcal E=\mathcal D\cap\mathbb R .
\label{eq:domain}
\end{equation}

\subsection{Conditions and theorem}
The criterion uses four conditions.
\begin{itemize}
\item[(C1)] Each $H_i$ is self-adjoint and bounded below.
\item[(C2)] $V$ is symmetric and $H_0$-bounded with relative bound $a<1$:
\begin{equation}
\|V\Psi\|\le a\,\|H_0\Psi\|+b\,\|\Psi\|,\qquad \Psi\in\dom(H_0).
\label{eq:relbound}
\end{equation}
\item[(C3)] For $E\in\mathcal D$, $\Sigma(E)$ extends from $\dom(H_1)$ to a bounded operator $\overline{\Sigma(E)}$ on $\mathcal H_1$.
\item[(C4)] There is an integer $m\ge1$ with $K(E)=R_1(E)\overline{\Sigma(E)}\in\SS m$ for $E\in\mathcal D$, and $\|K(E)\|_{\SS m}$ is bounded on compact subsets of $\mathcal D$.
\end{itemize}
The regularized determinant of order $m$ is~\cite{Simon2005}
\begin{equation}
{\det}_m(I-K)=\det\Bigl[(I-K)\exp\Bigl(\sum_{j=1}^{m-1}\frac{K^j}{j}\Bigr)\Bigr],\qquad K\in\SS m ,
\label{eq:detm}
\end{equation}
and ${\det}_m(I-K)=0$ exactly when $I-K$ is singular.

\begin{theorem}
\label{thm:main}
Assume {\rm (C1)--(C4)}. For every $E\in\mathcal D$,
\begin{equation}
E\in\sigma(H)\iff{\det}_m\bigl[I-K(E)\bigr]=0,
\qquad
\dim\ker(H-E)=\dim\ker\bigl[I-K(E)\bigr].
\label{eq:criterion}
\end{equation}
\end{theorem}

\begin{proof}
\emph{Step 1: domains.} By (C1), (C2), and the Kato--Rellich theorem~\cite{KatoBook},
\begin{equation}
H=H\dg,\qquad \dom(H)=\dom(H_0)=\bigoplus_i\dom(H_i).
\label{eq:domH}
\end{equation}
Let $P_{\rm c}$ be the projection onto $\bigoplus_{i\ge2}\mathcal H_i$ and set $H_0^{\rm c}=\operatorname{diag}(H_2,\dots,H_r)$. The internal complementary coupling is $V_{\rm c}=P_{\rm c}VP_{\rm c}$. For $\chi\in\dom(H_0^{\rm c})$,
\begin{equation}
\|V_{\rm c}\chi\|\le a\|H_0^{\rm c}\chi\|+b\|\chi\|.
\end{equation}
Hence $\Hc=H_0^{\rm c}+V_{\rm c}$ is self-adjoint on $\dom(H_0^{\rm c})$. Moreover,
\begin{equation}
(1-a)\|H_0^{\rm c}\chi\|\le \|\Hc\chi\|+b\|\chi\|,
\label{eq:graphnormc}
\end{equation}
so the graph norms of $H_0^{\rm c}$ and $\Hc$ are equivalent. Since $W=P_1VP_{\rm c}$ satisfies the same relative bound, for $E\in\mathcal D$ we obtain
\begin{equation}
B(E)=W(\Hc-E)^{-1}\in\mathcal B(\mathcal H_{\rm c},\mathcal H_1),
\qquad
C(E)=B(\bar E)\dg\in\mathcal B(\mathcal H_1,\mathcal H_{\rm c}).
\label{eq:BC}
\end{equation}
For $\phi\in\dom(H_1)$,
\begin{equation}
C(E)\phi=(\Hc-E)^{-1}W\dg\phi\in\dom(\Hc).
\label{eq:Cdomain}
\end{equation}

\emph{Step 2: Frobenius--Schur factorization.} Define
\begin{equation}
S(E)=H_1-E-\Sigma(E),\qquad \dom S(E)=\dom(H_1).
\end{equation}
On $\dom(H)$,
\begin{equation}
H-E=
\begin{pmatrix}I&B(E)\\0&I\end{pmatrix}
\begin{pmatrix}S(E)&0\\0&\Hc-E\end{pmatrix}
\begin{pmatrix}I&0\\C(E)&I\end{pmatrix}.
\label{eq:FS}
\end{equation}
The left factor is bounded and invertible on $\mathcal H$. The right factor is bounded and invertible on $\mathcal H$ and, by \eqref{eq:Cdomain}, maps $\dom(H)$ onto $\dom(H)$. Since $E\in\rho(\Hc)$, the factorization gives
\begin{equation}
E\in\sigma(H)\iff S(E)\text{ is singular},
\label{eq:SEa}
\end{equation}
and
\begin{equation}
\ker(H-E)=\bigl\{\phi\oplus[-C(E)\phi]:\phi\in\ker S(E)\bigr\}.
\label{eq:kerFS}
\end{equation}
Thus
\begin{equation}
\dim\ker(H-E)=\dim\ker S(E).
\label{eq:SE}
\end{equation}

\emph{Step 3: from $S(E)$ to $I-K(E)$.} On $\dom(H_1)$ we have
\begin{equation}
S(E)=(H_1-E)\bigl[I-K(E)\bigr],
\label{eq:SK}
\end{equation}
since $(H_1-E)R_1(E)\overline{\Sigma(E)}\phi=\overline{\Sigma(E)}\phi$. The range of $K(E)$ lies in $\ran R_1(E)=\dom(H_1)$, so every $\phi\in\ker[I-K(E)]$ lies in $\dom(H_1)$ and
\begin{equation}
\ker S(E)=\ker\bigl[I-K(E)\bigr].
\label{eq:kerSK}
\end{equation}
If $I-K(E)$ is invertible on $\mathcal H_1$, then for $\psi\in\dom(H_1)$
\begin{equation}
\bigl[I-K(E)\bigr]^{-1}\psi=\psi+K(E)\bigl[I-K(E)\bigr]^{-1}\psi\in\dom(H_1),
\end{equation}
so $I-K(E)$ maps $\dom(H_1)$ onto $\dom(H_1)$ and $S(E)$ is bijective. Conversely, $K(E)$ is compact by (C4), so $I-K(E)$ is invertible exactly when its kernel is $\{0\}$, and by \eqref{eq:kerSK} this holds when $S(E)$ is bijective.

\emph{Step 4: the determinant.} By (C4) and \eqref{eq:detm},
\begin{equation}
{\det}_m\bigl[I-K(E)\bigr]=0\iff\ker\bigl[I-K(E)\bigr]\neq\{0\}.
\end{equation}
Chaining with \eqref{eq:SE} and \eqref{eq:kerSK},
\begin{equation}
E\in\sigma(H)\iff\ker S(E)\neq\{0\}\iff\ker\bigl[I-K(E)\bigr]\neq\{0\}\iff{\det}_m\bigl[I-K(E)\bigr]=0,
\end{equation}
together with the equality of dimensions.
\end{proof}

\begin{remark}[Choice of the retained channel]
\label{rem:retained}
For two choices $i,j$ of the retained channel, Theorem~\ref{thm:main} gives the same zeros on $\mathcal E^{(i)}\cap\mathcal E^{(j)}$ and equal kernel dimensions. For $r=2$ the two kernels are $R_1V_{12}R_2V_{21}$ and $R_2V_{21}R_1V_{12}$, and for trace-class products
\begin{equation}
\det\bigl(I-AB\bigr)=\det\bigl(I-BA\bigr).
\end{equation}
The excluded set \eqref{eq:domain} and the index $m$ depend on the choice.
\end{remark}

\subsection{Full-space Birman--Schwinger form}
The same factorization can be applied before selecting a retained channel. This gives the Birman--Schwinger form used for the qubit arrays in Sec.~\ref{sec:qubits}.

\begin{proposition}
\label{prop:BS}
Assume {\rm (C1), (C2)}, let $E\in\mathbb R\setminus\sigma(H_0)$, $R_0(E)=(H_0-E)^{-1}$, and let $A(E)$ be the bounded extension of $R_0(E)V$. If $A(E)\in\SS m$, then
\begin{equation}
E\in\sigma(H)\iff{\det}_m\bigl[I+A(E)\bigr]=0,\qquad\dim\ker(H-E)=\dim\ker\bigl[I+A(E)\bigr].
\end{equation}
\end{proposition}
\begin{proof}
By \eqref{eq:relbound}, $VR_0(E)$ is bounded and $A(E)=[VR_0(E)]\dg$. On $\dom(H_0)$ we have
\begin{equation}
H-E=(H_0-E)\bigl[I+A(E)\bigr],
\end{equation}
and Steps 3 and 4 of the proof of Theorem~\ref{thm:main} apply with $H_1\to H_0$ and $K\to-A$.
\end{proof}

\section{Multiplicities}
\label{sec:mult}

This section shows that the order of a zero of ${\det}_m[I-K(E)]$ equals the multiplicity of the eigenvalue of $H$.

\begin{theorem}
\label{thm:mult}
Assume {\rm (C1)--(C4)} and let $E_0\in\mathcal E\cap\sigma(H)$. Then
\begin{equation}
\operatorname{ord}_{E_0}{\det}_m\bigl[I-K(E)\bigr]=\dim\ker(H-E_0).
\label{eq:ordmult}
\end{equation}
\end{theorem}

\begin{proof}
\emph{Step 1: reduction to $S(E)$.} By Lemma~\ref{lem:analytic} below, $E\mapsto K(E)$ is analytic in $\SS m$ near $E_0$. Writing ${\det}_m(I-K)=\det[(I-K)e^{Q}]$ with $Q=\sum_{j<m}K^j/j$, the factor $e^{Q(E)}$ is analytic and invertible, so the order of the zero equals the Gohberg--Sigal multiplicity of the family $I-K(E)$ at $E_0$~\cite{GohbergSigal,GGK2000}. By \eqref{eq:SK}, $I-K(E)=R_1(E)S(E)$, and $R_1(E)$ is analytic and invertible as a map from $\mathcal H_1$ onto $\dom(H_1)$ with the graph norm; hence the Gohberg--Sigal multiplicities of $I-K(E)$ and $S(E)$ at $E_0$ agree.

\emph{Step 2: sign of $S'(E_0)$.} For real $E$, $S(E)$ is symmetric on $\dom(H_1)$, and differentiating \eqref{eq:BC} gives
\begin{equation}
S'(E)=-I-\Sigma'(E),\qquad
\langle\phi,\Sigma'(E)\phi\rangle=\bigl\|(\Hc-E)^{-1}W\dg\phi\bigr\|^2\ge0 .
\end{equation}
Hence
\begin{equation}
\langle\phi,S'(E_0)\phi\rangle\le-\|\phi\|^2,\qquad\phi\in\dom(H_1).
\label{eq:Sprime}
\end{equation}

\emph{Step 3: Jordan chains.} A chain of length two at $E_0$ is a pair $\phi_0,\phi_1\in\dom(H_1)$ with
\begin{equation}
S(E_0)\phi_0=0,\qquad S(E_0)\phi_1+S'(E_0)\phi_0=0 .
\end{equation}
Taking the inner product of the second relation with $\phi_0$ and using the symmetry of $S(E_0)$,
\begin{equation}
\langle\phi_0,S'(E_0)\phi_0\rangle=-\langle S(E_0)\phi_0,\phi_1\rangle=0 .
\end{equation}
With \eqref{eq:Sprime} one obtains $\|\phi_0\|^2\le0$, so $\phi_0=0$. Every chain therefore has length one, and the Gohberg--Sigal multiplicity equals $\dim\ker S(E_0)$, which by \eqref{eq:SE} is $\dim\ker(H-E_0)$.
\end{proof}

\begin{remark}[Relation to the Weinstein--Aronszajn formula]
\label{rem:WA}
With $H_0'=H_1\oplus\Hc$ the Hamiltonian reads $H=H_0'+V'$, where the off-diagonal coupling $V'$ is built from $W$ and $W\dg$. For trace-class kernels the block identity
\begin{equation}
\det\begin{pmatrix} I & R_1(E)\,W\\ (\Hc-E)^{-1}W\dg & I\end{pmatrix}=\det\bigl[I-K(E)\bigr]
\end{equation}
turns the Birman--Schwinger determinant of the pair $(H_0',H)$ into the Schur determinant. Theorem~\ref{thm:mult} is then the instance on $\rho(H_1)\cap\rho(\Hc)$ of the global Weinstein--Aronszajn formula for modified Fredholm determinants~\cite[Theorem~5.4]{GLM2005}; see also~\cite{BehrndtGesztesy2020}, and~\cite{MuminovRasulov2015} for a multiplicity formula for the Schur complement of a $3\times3$ block operator matrix. The proof above gives a direct argument through the sign of $S'(E_0)$.
\end{remark}

\begin{corollary}
\label{cor:count}
Let $\Gamma$ be a closed contour on which ${\det}_m[I-K(E)]$ is zero-free, enclosing a region $\Omega$ with $\overline\Omega\subset\mathcal D$. Then
\begin{equation}
\frac{1}{2\pi i}\oint_\Gamma\frac{d}{dE}\log{\det}_m\bigl[I-K(E)\bigr]\,dE
=\sum_{E\in\Omega\cap\sigma(H)}\dim\ker(H-E).
\end{equation}
\end{corollary}

\section{Oscillator channels}
\label{sec:osc}

This section gives the matrix elements of $K(E)$ when channels are equivalent to harmonic oscillators. We call channel $i$ an oscillator channel if there is a unitary $U_i$ with
\begin{equation}
U_i\dg H_iU_i=\sum_{k=1}^{N_b}\Omega_{ik}\,\ad_ka_k+\varepsilon_i,\qquad\Omega_{ik}>0 ,
\label{eq:normalform}
\end{equation}
where $N_b$ is the number of modes. In the Fock basis $\ket{\mathbf n}$ of the right-hand side the resolvent is diagonal, and
\begin{equation}
\bra{\mathbf m}U_i\dg R_i(E)V_{ij}U_j\ket{\mathbf n}
=\frac{\bra{\mathbf m}U_i\dg V_{ij}U_j\ket{\mathbf n}}{\sum_k\Omega_{ik}m_k+\varepsilon_i-E}.
\label{eq:symbol}
\end{equation}
For a displacement $U=D(\alpha)$, $\alpha\in\mathbb R$, and $m\ge n$,
\begin{equation}
\bra mD(\alpha)\ket n=\sqrt{\frac{n!}{m!}}\,\alpha^{m-n}e^{-\alpha^2/2}L_n^{(m-n)}(\alpha^2),
\qquad
\bra nD(\alpha)\ket m=(-1)^{m-n}\bra mD(\alpha)\ket n ,
\label{eq:Dmatrix}
\end{equation}
with associated Laguerre polynomials $L_n^{(k)}$; squeezings give Hermite polynomials. For a channel with continuous spectrum the resolvent follows from the spectral theorem.

\section{Regularization order}
\label{sec:order}

This section determines the smallest $m$ in (C4) for oscillator channels.

\subsection{Upper bound}
The oscillator resolvent fixes the upper Schatten estimate.
\begin{lemma}
\label{lem:resolvent}
For $R(E)=\bigl(\sum_{k=1}^{N_b}\omega_k\ad_ka_k+\varepsilon-E\bigr)^{-1}$ with $\omega_k>0$,
\begin{equation}
R(E)\in\SS p\iff p>N_b .
\end{equation}
\end{lemma}
\begin{proof}
The eigenvalues are $\lambda_{\mathbf n}=(\sum_k\omega_kn_k+\varepsilon-E)^{-1}$, and the number of $\mathbf n$ with $\sum_k\omega_kn_k\le L$ grows as $L^{N_b}$. Hence
\begin{equation}
\sum_{\mathbf n}|\lambda_{\mathbf n}|^p\sim\int^\infty L^{N_b-1-p}\,dL ,
\end{equation}
finite exactly for $p>N_b$.
\end{proof}

\begin{proposition}
\label{prop:upper}
Let $K(E)$ be a product of two oscillator resolvents with $N_b$ modes and bounded operators. Then $K(E)\in\SS q$ for every $q>N_b/2$, and
\begin{equation}
m=\Bigl\lfloor\frac{N_b}{2}\Bigr\rfloor+1
\label{eq:mcount}
\end{equation}
satisfies {\rm (C4)}.
\end{proposition}
\begin{proof}
By Lemma~\ref{lem:resolvent} each resolvent lies in $\SS p$ for $p>N_b$, and the H\"older inequality $\|AB\|_{\SS q}\le\|A\|_{\SS p}\|B\|_{\SS{p'}}$ with $1/q=1/p+1/p'$~\cite{Simon2005} gives $q>N_b/2$. The integer \eqref{eq:mcount} is the smallest integer above $N_b/2$.
\end{proof}

For comparison, the full-space kernel $A(E)$ of Proposition~\ref{prop:BS} carries a single resolvent factor and lies in $\SS p$ for $p>N_b$. The Schur reduction therefore halves the Schatten index.

For couplings of the form $g(a+\ad)$ the operator between the two resolvents is unbounded, while Proposition~\ref{prop:upper} requires a bounded one. The following proposition supplies the upper bound in this case.
\begin{proposition}[Linear couplings]
\label{prop:upperlin}
Let channel $1$ be an oscillator channel with $N_b$ modes, let $\mathcal N_{\rm c}$ be the number operator of the complementary channels, and let $E_0<\min\{\inf\sigma(H_1),\inf\sigma(\Hc)\}$. Assume that, with constants $c,C>0$,
\begin{equation}
\Hc-E_0\ge c\,(\mathcal N_{\rm c}+1),\qquad
\bigl\|(\mathcal N_{\rm c}+1)^{-1/2}W\dg\phi\bigr\|\le C\,\|\phi\|,\quad\phi\in\dom(H_1),
\label{eq:lincond}
\end{equation}
the first inequality in the form sense. Then {\rm (C3)} holds, $K(E)\in\SS p$ for every $p>N_b$ and $E\in\mathcal D$, and $\|K(E)\|_{\SS p}$ is bounded on compact subsets of $\mathcal D$. In particular {\rm (C4)} holds with $m=N_b+1$, and with $m=2$ for a single-mode channel.
\end{proposition}
\begin{proof}
The proof factorizes the self-energy through a bounded operator. The operator monotonicity of the inverse and the first inequality in \eqref{eq:lincond} give
\begin{equation}
(\Hc-E_0)^{-1}\le c^{-1}(\mathcal N_{\rm c}+1)^{-1}.
\end{equation}
Hence, for $\phi\in\dom(H_1)$,
\begin{equation}
\bigl\|(\Hc-E_0)^{-1/2}W\dg\phi\bigr\|^2
=\bigl\langle W\dg\phi,(\Hc-E_0)^{-1}W\dg\phi\bigr\rangle
\le c^{-1}\bigl\|(\mathcal N_{\rm c}+1)^{-1/2}W\dg\phi\bigr\|^2
\le\frac{C^2}{c}\,\|\phi\|^2 ,
\end{equation}
so $(\Hc-E_0)^{-1/2}W\dg$ extends to a bounded operator $B\dg$ with $\|B\|\le C/\sqrt c$. For $E\in\mathcal D$ the functional calculus gives the bounded operators
\begin{equation}
T_{\rm c}(E)=(\Hc-E_0)(\Hc-E)^{-1},\qquad T_1(E)=(H_1-E_0)(H_1-E)^{-1},
\end{equation}
and on $\dom(H_1)$
\begin{equation}
\Sigma(E)=B\,T_{\rm c}(E)\,B\dg,\qquad
K(E)=T_1(E)\,R_1(E_0)\,B\,T_{\rm c}(E)\,B\dg .
\end{equation}
The first identity exhibits a bounded extension of $\Sigma(E)$, which is {\rm (C3)}. By Lemma~\ref{lem:resolvent}, $R_1(E_0)\in\SS p$ for $p>N_b$, and the ideal property gives
\begin{equation}
\|K(E)\|_{\SS p}\le\|T_1(E)\|\,\|R_1(E_0)\|_{\SS p}\,\frac{C^2}{c}\,\|T_{\rm c}(E)\| .
\end{equation}
The norms of $T_1(E)$ and $T_{\rm c}(E)$ are continuous on $\mathcal D$, so the bound is uniform on compact subsets.
\end{proof}

For linear couplings the conditions \eqref{eq:lincond} follow from two elementary estimates. With $a a\dg=\mathcal N+1$ and $\ad a\le\mathcal N+1$,
\begin{equation}
\bigl\|(\mathcal N+1)^{-1/2}(a+\ad)\psi\bigr\|\le2\,\|\psi\| ,
\end{equation}
which gives the second condition with $C=2\max|g|$. With $\|a\psi\|^2=\langle\psi,\mathcal N\psi\rangle$, $\|\ad\psi\|^2=\langle\psi,(\mathcal N+1)\psi\rangle$ and the Cauchy--Schwarz inequality, an internal coupling $V_{\rm c}$ of $\Hc$ built from $g'(a+\ad)$ and bounded spin factors obeys
\begin{equation}
|\langle\psi,V_{\rm c}\psi\rangle|\le\delta\,\langle\psi,\mathcal N_{\rm c}\psi\rangle+C_\delta\,\|\psi\|^2
\end{equation}
for every $\delta>0$. The internal couplings are therefore form-bounded with relative bound zero with respect to the oscillator part of $\Hc$, and the first condition holds for $E_0$ sufficiently negative. The Rabi--Stark model in the $\sigma_z$ decomposition, the three-level emitter and the Dicke model have this structure.

\subsection{Lower bounds}
Two inequalities give lower bounds. For $K\in\SS p$, $p\ge1$, and any orthonormal basis $\{e_n\}$~\cite{Simon2005},
\begin{equation}
\sum_n\bigl|\langle e_n,Ke_n\rangle\bigr|^p\le\|K\|_{\SS p}^p .
\label{eq:diagbound}
\end{equation}
For $A>0$ and $\psi\in\dom(A^{1/2})$, the Cauchy--Schwarz inequality $\|\psi\|^2=\langle A^{1/2}\psi,A^{-1/2}\psi\rangle\le\|A^{1/2}\psi\|\,\|A^{-1/2}\psi\|$ gives
\begin{equation}
\langle\psi,A^{-1}\psi\rangle\ge\frac{\|\psi\|^4}{\langle\psi,A\psi\rangle}.
\label{eq:jensen}
\end{equation}

\begin{proposition}[Bounded couplings]
\label{prop:lowerb}
Let $r=2$ with oscillator channels of $N_b$ modes, total number operator $\mathcal N$, and coupling $U_1\dg V_{12}U_2=c\,U$ with $c\neq0$ and $U$ unitary satisfying
\begin{equation}
U\dg(\mathcal N+1)U\le C_U(\mathcal N+1)
\label{eq:growth}
\end{equation}
in the form sense. Then $\|K(E)\|_{\SS p}=\infty$ for $p\le N_b/2$, and \eqref{eq:mcount} is the smallest integer in {\rm (C4)}.
\end{proposition}
\begin{proof}
Take $E<\inf\sigma(H_0)$, so that $\tilde R_i=U_i\dg R_iU_i>0$, and write $r_{\mathbf n}$ for the eigenvalues of $\tilde R_1$. In the Fock basis of the normal form,
\begin{equation}
\bra{\mathbf n}U_1\dg KU_1\ket{\mathbf n}=c^2\,r_{\mathbf n}\,\bra{U\dg\mathbf n}\tilde R_2\ket{U\dg\mathbf n}.
\end{equation}
By \eqref{eq:jensen} and \eqref{eq:growth},
\begin{equation}
\bra{U\dg\mathbf n}\tilde R_2\ket{U\dg\mathbf n}\ge\frac{1}{\bra{U\dg\mathbf n}\tilde R_2^{-1}\ket{U\dg\mathbf n}}\ge\frac{c_1}{1+|\mathbf n|},
\qquad r_{\mathbf n}\ge\frac{c_2}{1+|\mathbf n|},
\end{equation}
with $|\mathbf n|=\sum_kn_k$. The number of $\mathbf n$ with $|\mathbf n|=L$ is $\binom{L+N_b-1}{N_b-1}\ge L^{N_b-1}/(N_b-1)!$. With \eqref{eq:diagbound} one obtains
\begin{equation}
\|K\|_{\SS p}^p\ge c_3\sum_{L\ge1}L^{N_b-1}(1+L)^{-2p}=\infty\qquad(2p\le N_b).
\end{equation}
Condition (C4) asks for one $m$ valid on all of $\mathcal D$, so the value at one $E$ fixes the smallest admissible integer.
\end{proof}

Displacements and squeezings satisfy \eqref{eq:growth}.

\begin{proposition}[Unbounded couplings]
\label{prop:lowerunb}
Let channel $1$ be a single-mode oscillator with $\tilde R_1$ eigenvalues $r_n\ge c_1/(1+n)$, and let $v_n=W\dg U_1\ket n$ satisfy, for some $E<\inf\sigma(H)$,
\begin{equation}
\|v_n\|^2\ge c_2\,(1+n),\qquad
\langle v_n,(\Hc-E)v_n\rangle\le c_3\,(1+n)\,\|v_n\|^2 .
\label{eq:unbcond}
\end{equation}
Then $\bra nU_1\dg KU_1\ket n\ge c/(1+n)$, $K(E)\notin\SS1$, and $m\ge2$.
\end{proposition}
\begin{proof}
By \eqref{eq:jensen} and \eqref{eq:unbcond},
\begin{equation}
\bra nU_1\dg KU_1\ket n=r_n\,\langle v_n,(\Hc-E)^{-1}v_n\rangle
\ge r_n\frac{\|v_n\|^4}{\langle v_n,(\Hc-E)v_n\rangle}
\ge\frac{c_1c_2}{c_3}\,\frac{1}{1+n},
\end{equation}
and \eqref{eq:diagbound} with $p=1$ gives $\|K\|_{\SS1}=\infty$.
\end{proof}

For couplings $g(a+\ad)$ between oscillator channels the conditions \eqref{eq:unbcond} hold whenever $v_n$ lies in channels whose mutual couplings vanish on $v_n$, as in Secs.~\ref{sec:jc}, \ref{sec:lambda} and~\ref{sec:dicke}; then $\langle v_n,\Hc v_n\rangle$ is the diagonal expectation value, of order $(1+n)\|v_n\|^2$. Proposition~\ref{prop:upperlin} gives the matching upper bound, so $m=2$ for these models.

\section{Analytic structure and truncation}
\label{sec:analytic}

This section establishes analyticity in $E$, describes the singularities of ${\det}_m$, and proves convergence of truncated determinants.

\subsection{Analyticity}
We first establish analytic dependence on the spectral parameter.
\begin{lemma}
\label{lem:analytic}
Under {\rm (C1)--(C4)}, $E\mapsto K(E)$ and $E\mapsto{\det}_m[I-K(E)]$ are analytic on each component of $\mathcal D$.
\end{lemma}
\begin{proof}
For $E,E_0\in\mathcal D$ the resolvent identity gives
\begin{equation}
\Sigma(E)-\Sigma(E_0)=(E-E_0)\,B(E)\,C(E_0),
\end{equation}
with $B$, $C$ of \eqref{eq:BC} bounded and norm-continuous, so $K(E)$ is norm-analytic. For finite-rank $F$ the functions $E\mapsto\tr[FK(E)]$ are analytic and separate points of $\SS m$; with the local bound of (C4), the theorem of Arendt and Nikolski~\cite{ArendtNikolski2000} gives analyticity in $\SS m$. The map $A\mapsto{\det}_m(I-A)$ is analytic on $\SS m$~\cite{Simon2005}.
\end{proof}

\subsection{Singularities}
The determinant is analytic on $\mathcal D$. Its boundary points inherit the singular structure of the channel resolvents. Suppose that an isolated point $E_\ast\in\sigma(H_1)\cup\sigma(\Hc)$ is an eigenvalue of finite multiplicity and that the corresponding resolvent has a first-order pole. Then $K(E)$ has a Laurent expansion
\begin{equation}
K(E)=\frac{F}{E-E_\ast}+K_{\rm reg}(E),
\end{equation}
with finite-rank $F$. For $m=1$ the Fredholm determinant is meromorphic. For $m\ge2$ the exponential regularization can generate an essential singularity. The scalar identity
\begin{equation}
{\det}_2(1-k)=(1-k)e^k,\qquad k=\frac{c}{E-E_\ast},
\end{equation}
displays this behavior directly. Numerical root searches are therefore performed on compact subsets of $\mathcal D$.

When a finite-rank approximation $A_N$ of an operator $A$ is available, the following lemma gives meromorphy of suitably renormalized determinants.

\begin{lemma}
\label{lem:mero}
Let $A(E)\in\SS2$ be analytic on $B\setminus\{E_\ast\}$, $B$ a disc around $E_\ast$, and let $A_N(E)$ be finite-rank with $A_N\to A$ in $\SS2$ and $\tr A_N\to t$ uniformly on circles in $B\setminus\{E_\ast\}$. If $(E-E_\ast)^q\det[I+A_N(E)]$ is analytic in $B$ for every $N$, then
\begin{equation}
f(E)={\det}_2\bigl[I+A(E)\bigr]\,e^{t(E)}
\end{equation}
satisfies: $(E-E_\ast)^qf(E)$ extends analytically to $B$.
\end{lemma}
\begin{proof}
For finite rank, $\det(I+A_N)={\det}_2(I+A_N)\,e^{\tr A_N}$, which converges to $f$ uniformly on a circle $|E-E_\ast|=\rho$. The functions $(E-E_\ast)^q\det(I+A_N)$ are analytic in the disc, and by the maximum principle they converge uniformly on the closed disc. The limit is analytic and equals $(E-E_\ast)^qf$ on the punctured disc.
\end{proof}

\subsection{Truncation}
We next pass from the infinite kernel to finite Fock projections.
\begin{proposition}
\label{prop:trunc}
Assume {\rm (C1)--(C4)} with oscillator channels, let $P_N$ project onto the first $N$ Fock states of the normal form, and $K_N=P_NKP_N$. On every compact $\mathcal C$ in a component of $\mathcal D$,
\begin{equation}
\sup_{E\in\mathcal C}\|K(E)-K_N(E)\|_{\SS m}\to0,\qquad
\sup_{E\in\mathcal C}\bigl|{\det}_m(I-K_N)-{\det}_m(I-K)\bigr|\to0 .
\label{eq:truncconv}
\end{equation}
For every disc $B$ with $\overline B\subset\mathcal C$ and ${\det}_m(I-K)$ zero-free on $\partial B$, the number of zeros in $B$, counted with order, is the same for ${\det}_m(I-K_N)$ and ${\det}_m(I-K)$ for large $N$. Every limit point in $\mathcal C$ of zeros of ${\det}_m(I-K_N)$ is a zero of ${\det}_m(I-K)$.
\end{proposition}
\begin{proof}
We have
\begin{equation}
K-P_NKP_N=(I-P_N)K+P_NK(I-P_N),
\end{equation}
and both terms tend to zero in $\SS m$ by density of finite-rank operators in $\SS m$; uniformity on $\mathcal C$ follows from Lemma~\ref{lem:analytic}. The bound~\cite{Simon2005}
\begin{equation}
\bigl|{\det}_m(I-A)-{\det}_m(I-B)\bigr|\le\|A-B\|_{\SS m}\exp\bigl[\Gamma_m\bigl(\|A\|_{\SS m}+\|B\|_{\SS m}+1\bigr)^m\bigr],
\end{equation}
with a constant $\Gamma_m$ depending on $m$, gives the second limit. Uniform convergence on $\partial B$ and Rouch\'e's theorem give the zero count; the statement on limit points follows from local uniform convergence.
\end{proof}

In computations each factor of $K$ is truncated separately. We use
\begin{equation}
K'_N=c^2(P_NR_1P_N)(P_NUP_N)(P_NR_2P_N)(P_NU\dg P_N).
\end{equation}
Writing $K-K'_N$ as a telescoping sum places $I-P_N$ next to a resolvent factor. Since
\begin{equation}
\|(I-P_N)R_i\|_{\SS2}\longrightarrow0,
\end{equation}
one obtains $K'_N\to K$ in $\SS1$ for one mode.

For a rate, take two one-mode channels and define
\begin{equation}
\mathcal N=\ad a,\qquad
A_{12}=cR_1U,\qquad
A_{21}=cR_2U\dg,
\end{equation}
with
\begin{equation}
K=A_{12}A_{21},\qquad
R_i=(\Omega\mathcal N+\varepsilon_i-E)^{-1},\qquad
U=D(\alpha),\quad \alpha\in\mathbb R.
\end{equation}
Then
\begin{equation}
\|(I-P_N)K\|_{\SS1}\le\|(I-P_N)R_1\|_{\SS2}\|A_{21}\|_{\SS2}\le\frac{c_1}{\sqrt N},
\qquad
\|A_{21}(I-P_N)\|_{\SS2}^2=c^2\sum_{n\ge N}\bra n UR_2^2U\dg\ket n .
\end{equation}
With $p_{mn}=|\bra mD(\alpha)\ket n|^2$ and $f(m)=|\Omega m+\varepsilon_2-E|^{-2}\le c_2(1+m)^{-2}$, split the sum over $m$ at $m=n/2$. For $m\ge n/2$ one obtains $f(m)\le4c_2(1+n)^{-2}$. For $m<n/2$ we have $|m-n|>n/2$ and
\begin{equation}
\sum_{m<n/2}p_{mn}\le\Bigl(\frac2n\Bigr)^4\sum_m(m-n)^4p_{mn}
=\Bigl(\frac2n\Bigr)^4\bigl\|(\mathcal N-n)^2D(\alpha)\ket n\bigr\|^2 .
\end{equation}
With $Y=D(\alpha)\dg\mathcal ND(\alpha)-\mathcal N=\alpha(a+\ad)+\alpha^2$ and $(\mathcal N-n)\ket n=0$,
\begin{equation}
D(\alpha)\dg(\mathcal N-n)^2D(\alpha)\ket n=(\mathcal N-n+Y)^2\ket n=(\mathcal N-n)Y\ket n+Y^2\ket n ,
\end{equation}
\begin{equation}
\bigl\|(\mathcal N-n)Y\ket n\bigr\|^2=\alpha^2(2n+1),\qquad
\bigl\|Y^2\ket n\bigr\|^2=O(\alpha^4n^2).
\end{equation}
Hence $\sum_{m<n/2}p_{mn}=O(\alpha^4n^{-2})$, $\bra nUR_2^2U\dg\ket n=O(n^{-2})$, and
\begin{equation}
\sup_{E\in\mathcal C}\|K(E)-K_N(E)\|_{\SS1}\le\frac{C(\mathcal C)}{\sqrt N}.
\label{eq:rate}
\end{equation}

\section{Models}
\label{sec:models}

This section applies Theorems~\ref{thm:main}--\ref{thm:mult} and Secs.~\ref{sec:osc}--\ref{sec:analytic} to nine models. Table~\ref{tab:models} collects the decompositions, normal-form frequencies and indices.

\begin{table}[t]
\centering
\small
\begin{tabular}{@{}lllcl@{}}
\toprule
Model & Decomposition & $\Omega$ & $m$ & Boundary of the class\\
\midrule
Friedrichs--Lee & discrete--continuum & --- & 1 & --- \\
Quantum Rabi & $\sigma_x$ & $\omega$ & 1 & --- \\
Jaynes--Cummings & $\sigma_z$ & $\omega$ & 2 & --- \\
Rabi--Stark & $\sigma_z$ & $\omega\pm\chi$ & 2 & $|\chi|=\omega$ \\
Two-photon & $\sigma_x$ & $\sqrt{\omega^2-4g^2}$ & 1 & $g=\omega/2$ \\
Two-mode & $\sigma_x$ & $\sqrt{\omega^2-g^2}$ & 2 & $g=\omega$ \\
Three-level $\Lambda$ & atomic levels & $\omega$ & 2 & --- \\
Dicke & $J_z$ & $\omega$ & 2 & --- \\
Qubit array & $\otimes_i\sigma_{i,x}$ & $\omega$ & $1^{a}$ & --- \\
\bottomrule
\end{tabular}
\caption{Decomposition, normal-form frequency $\Omega$ and smallest index $m$ in (C4) for each model; every listed $m$ carries a matching lower bound from Propositions~\ref{prop:lowerb} or~\ref{prop:lowerunb}. The last column gives the coupling at which a channel loses its oscillator normal form. $^{a}$Schur reduction to one channel; the parity-reduced determinant of Sec.~\ref{sec:qubits} is a ${\det}_2$ with vanishing trace.}
\label{tab:models}
\end{table}

\subsection{Friedrichs--Lee model}
\label{sec:FL}
\begin{figure}[t]
\centering
\includegraphics[width=0.6\textwidth]{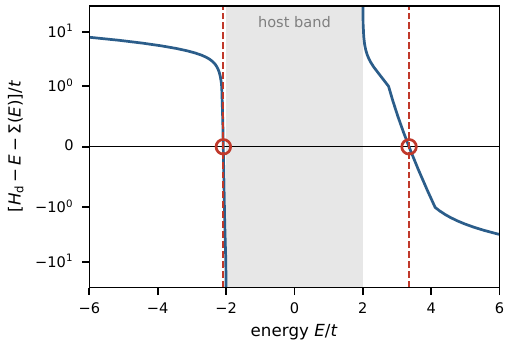}
\caption{Impurity level in a tight-binding band of half-bandwidth $2t$ ($\varepsilon_d=2.4t$, $V=1.6t$, $600$ band points). Solid curve: $H_{\rm d}-E-\Sigma(E)$ on a symmetric-logarithmic scale; circles: its zeros; dashed lines: bound levels from direct diagonalization. The shaded region is the band.}
\label{fig:impurity}
\end{figure}

This model has a channel with continuous spectrum and a kernel of finite rank. With a $d\times d$ Hermitian matrix $H_{\rm d}$, a multiplication operator $\Hc$ with spectral variable $k$ and dispersion $\omega(k)$, and coupling $V$~\cite{Friedrichs1948,Lee1954},
\begin{equation}
H=\begin{pmatrix}H_{\rm d}&V\dg\\ V&\Hc\end{pmatrix},\qquad
K(E)=(H_{\rm d}-E)^{-1}V\dg(\Hc-E)^{-1}V .
\end{equation}
The kernel has rank at most $d$, so $m=1$, and
\begin{equation}
\det\bigl[I-K(E)\bigr]=\frac{\det\bigl[H_{\rm d}-E-\Sigma(E)\bigr]}{\det\bigl[H_{\rm d}-E\bigr]} .
\end{equation}
Theorem~\ref{thm:main} gives the discrete spectrum outside $\sigma(\Hc)$ as the zeros of
\begin{equation}
\det\bigl[H_{\rm d}-E-\Sigma(E)\bigr]=0 .
\label{eq:FL}
\end{equation}
For $d=1$, $H_{\rm d}=\varepsilon_d$ and form factor $g(k)$, one obtains the level-shift equation of the Friedrichs--Lee and Fano models~\cite{KosterSlater1954,Fano1961},
\begin{equation}
E=\varepsilon_d+\int dk\,\frac{|g(k)|^2}{E-\omega(k)} .
\end{equation}
For a tight-binding band of half-bandwidth $2t$ discretized on $600$ points, with $\varepsilon_d=2.4t$ and $V=1.6t$, the root of \eqref{eq:FL} and the bound state from direct diagonalization agree to $3\times10^{-15}$ (Fig.~\ref{fig:impurity}).

\subsection{Quantum Rabi model}
\label{sec:rabi}
This subsection applies the general construction to the quantum Rabi model and then resolves the determinant into its two parity sectors. The Hamiltonian is
\begin{equation}
H=\omega\ad a+\frac{\Delta}{2}\sigma_z+g\,(a+\ad)\,\sigma_x ,
\label{eq:rabi}
\end{equation}
where $\omega$ is the mode frequency, $\Delta$ the qubit splitting, and $g$ the coupling. In the eigenbasis $\ket\pm$ of $\sigma_x$ we write
\begin{equation}
\Psi=\phi_+\ket+ +\phi_-\ket- .
\end{equation}
The eigenvalue equation gives
\begin{equation}
\begin{aligned}
(H_+-E)\phi_++\frac{\Delta}{2}\phi_-&=0,\\
\frac{\Delta}{2}\phi_++(H_--E)\phi_-&=0,
\end{aligned}
\qquad
H_\pm=\omega\ad a\pm g(a+\ad).
\label{eq:rabirows}
\end{equation}
With $R_\pm(E)=(H_\pm-E)^{-1}$, the second row gives
\begin{equation}
\phi_-=-\frac{\Delta}{2}R_-(E)\phi_+ .
\end{equation}
Substitution into the first row yields
\begin{equation}
\bigl[I-K(E)\bigr]\phi_+=0,
\qquad
K(E)=\frac{\Delta^2}{4}R_+(E)R_-(E).
\label{eq:rabiK}
\end{equation}
Set $\lambda=g/\omega$ and $D(\alpha)=e^{\alpha(\ad-a)}$. We have
\begin{equation}
D(\mp\lambda)\dg H_\pm D(\mp\lambda)
=\omega\ad a-\frac{g^2}{\omega}.
\end{equation}
Thus both channels reduce to the same oscillator resolvent
\begin{equation}
R(E)=\left(\omega\ad a-\frac{g^2}{\omega}-E\right)^{-1},
\end{equation}
and
\begin{equation}
D(\lambda)K(E)D(\lambda)\dg
=\frac{\Delta^2}{4}R(E)D(2\lambda)R(E)D(-2\lambda).
\label{eq:rabiKtilde}
\end{equation}
In the Fock basis, the singular values of $R(E)$ scale as $n^{-1}$. Hence $R(E)\in\SS2$, and the H\"older inequality gives $K(E)\in\SS1$.

\begin{figure}[t]
\centering
\includegraphics[width=0.6\textwidth]{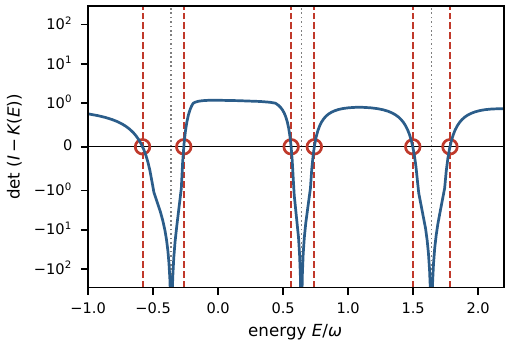}
\caption{Quantum Rabi model at $g/\omega=0.6$, $\Delta/\omega=0.7$, truncation $N=80$. Solid curve: $\det[I-K(E)]$ on a symmetric-logarithmic scale; circles: zeros; dashed lines: eigenvalues from direct diagonalization, agreeing to $7\times10^{-15}$; dotted lines: the ladder $\sigma(H_\pm)$.}
\label{fig:rabi}
\end{figure}

\emph{Conditions.} The channels $H_\pm$ are displaced oscillators with $\Omega=\omega$, and the coupling is the scalar $\Delta/2$; (C1)--(C3) hold, and Eq.~\eqref{eq:rabiKtilde} gives (C4) with $m=1$. The excluded set is the common ladder
\begin{equation}
\sigma(H_+)=\sigma(H_-)=\Bigl\{n\omega-\frac{g^2}{\omega}:\ n\in\mathbb N_0\Bigr\}.
\end{equation}
The determinant and its zeros are shown in Fig.~\ref{fig:rabi}. At $\omega=\Delta=g=1$, the zeros of $\det[I-K_N(E)]$ at $N=60$ on $(-1.6,3.5)$ are
\begin{equation}
\begin{aligned}
&-1.147945729,\ -1.010178301,\ -0.231722500,\ 0.133435448,\ 0.927043874,\\
&\phantom{-}1.104809460,\ \phantom{-}1.842780994,\ \phantom{-}2.143619454,\ 2.943927717,\ 3.043187919,
\end{aligned}
\end{equation}
equal to the ten lowest eigenvalues from direct diagonalization at Fock cutoff $900$.

\emph{Exceptional points.} On the ladder the elimination is replaced by a residue condition. Writing $\phi_\pm=c_\pm\ket{n_\ast}+\phi_\pm^\perp$ with $\ket{n_\ast}$ spanning $\ker(H_\pm-E)$, the component along $\ket{n_\ast}$ is a finite-rank condition on the parameters, which reproduces the Judd solutions~\cite{Judd1979}; at $\Delta=2N\omega$, $N=1,2$, its root agrees with the exceptional eigenvalue to thirteen digits.

\emph{Parity factorization.} With $\Pi=(-1)^{\ad a}$ we have
\begin{equation}
\Pi D(\alpha)\Pi=D(-\alpha),\qquad [\Pi,R(E)]=0 .
\end{equation}
Hence the kernel \eqref{eq:rabiKtilde} is a square,
\begin{equation}
D(\lambda)K(E)D(\lambda)\dg=X(E)^2,\qquad
X(E)=\frac{\Delta}{2}\,R(E)\,D(2\lambda)\,\Pi .
\label{eq:X}
\end{equation}
Here $X(E)\in\SS2$, with diagonal entries
\begin{equation}
X_{nn}(E)=\frac{\Delta}{2}\,\frac{(-1)^n e^{-2\lambda^2}L_n(4\lambda^2)}{n\omega-g^2/\omega-E}.
\end{equation}
For fixed $y>0$ the Fej\'er asymptotics~\cite{Szego1975}
\begin{equation}
e^{-y/2}L_n(y)=\frac{\cos\bigl(2\sqrt{ny}-\pi/4\bigr)}{\sqrt\pi\,(ny)^{1/4}}+O\bigl(n^{-3/4}\bigr)
\label{eq:fejer}
\end{equation}
give $X_{nn}=O(n^{-5/4})$ uniformly on compact sets away from the ladder, so
\begin{equation}
\tau(E)=\sum_{n\ge0}X_{nn}(E)
\end{equation}
converges absolutely and is meromorphic with simple poles on the ladder. We define the parity determinants
\begin{equation}
D_\pm(E)={\det}_2\bigl[I\mp X(E)\bigr]\,e^{\mp\tau(E)} .
\label{eq:Dpm}
\end{equation}
For finite rank we have ${\det}_2(I\mp X_N)e^{\mp\tr X_N}=\det(I\mp X_N)$, and $X_N$ has a pole in one row at each ladder point, so Lemma~\ref{lem:mero} with $q=1$ shows that $D_\pm$ are meromorphic with at most simple poles on the ladder. With $(I-X)e^{X}(I+X)e^{-X}=I-X^2$,
\begin{equation}
D_+(E)\,D_-(E)=\det\bigl[I-X(E)^2\bigr]=\det\bigl[I-K(E)\bigr] .
\label{eq:Dproduct}
\end{equation}
The zeros of $D_+$ and $D_-$ split the spectrum on $\mathcal E$ into the two eigenspaces of the parity $\sigma_z(-1)^{\ad a}$.

\emph{Computational cost.} The determinant representation has a numerical cost, measured here by the Fock cutoff. Set $\omega=\Delta=1$. The smallest Fock cutoff that reproduces the ground-state energy to $10^{-8}$ is
\begin{equation}
\begin{array}{lccccc}
g/\omega & 0.7 & 1.0 & 1.5 & 2.0 & 3.0\\
\text{direct diagonalization} & 10 & 12 & 18 & 22 & 34\\
\text{root of }\det(I-K_N) & 12 & 18 & 28 & 40 & 68
\end{array}
\end{equation}
The kernel \eqref{eq:rabiKtilde} contains $D(\pm2\lambda)$, a displacement by twice the polaron shift, so the truncation accommodates a spread of order $4g^2/\omega^2$. The determinant supplies an analytic function of $E$: by Corollary~\ref{cor:count} the eigenvalues in a region are counted with multiplicity from a contour integral, and for a continuum channel the Friedrichs--Lee reduction \eqref{eq:FL} is exact at finite dimension $d$.

\emph{Relation to Braak's $G$-functions.} Set $\omega=1$, so that $\lambda=g$, and write $x=E+g^2$. Braak's functions~\cite{Braak2011} for $\Delta_{\rm B}=\Delta/2$ are
\begin{equation}
G_\pm(x)=\sum_{n\ge0}k_n(x)\Bigl[1\mp\frac{\Delta_{\rm B}}{x-n}\Bigr]g^n,
\label{eq:braakG}
\end{equation}
\begin{equation}
k_0=1,\quad k_1=f_0,\quad nk_n=f_{n-1}k_{n-1}-k_{n-2},\quad
f_n(x)=2g+\frac{1}{2g}\Bigl[n-x+\frac{\Delta_{\rm B}^2}{x-n}\Bigr].
\end{equation}

\begin{conjecture}
\label{conj:G}
For $\omega=1$,
\begin{equation}
D_\pm(E)=e^{-2g^2}\,2^{x}\,G_\mp(x),\qquad x=E+g^2,
\label{eq:conjG}
\end{equation}
and therefore, by \eqref{eq:Dproduct},
\begin{equation}
\det\bigl[I-K(E)\bigr]=e^{-4g^2}\,4^{x}\,G_+(x)\,G_-(x).
\label{eq:conjGprod}
\end{equation}
\end{conjecture}

The numerical evidence is the following. At $g=0.7$, $\Delta=1$, the zeros of $D_+$ coincide with those of $G_-$, and the zeros of $D_-$ with those of $G_+$, for the eight lowest levels. The ratios $D_+(E)/[e^{-2g^2}2^xG_-(x)]$ and $D_-(E)/[e^{-2g^2}2^xG_+(x)]$, listed in this order, are constant in $E$ to four digits and approach one with the truncation:
\begin{equation}
\begin{array}{lccc}
(g,\Delta) & (0.7,1.0) & (0.4,1.6) & (1.1,0.6)\\
N=120 & 1.0011,\ 1.0011 & 1.0025,\ 1.0031 & 1.0003,\ 1.0006\\
N=240 & 1.0006,\ 1.0005 & 1.0015,\ 1.0013 & 1.0002,\ 1.0002
\end{array}
\end{equation}
The leading truncation error of ${\det}_2$ is $\tfrac12\tr\bigl(X^2-X_N^2\bigr)$, which for $\omega=1$ gives
\begin{equation}
\frac{\Delta^2}{8}\sum_{n\ge N}\frac{1}{(n-x)^2}\simeq\frac{\Delta^2}{8N};
\end{equation}
this term accounts for the listed deviations from one.

For the product, $\log\{\det[I-K(E)]/[G_+(x)G_-(x)]\}$ is linear on $E\in(-1.5,3.6)$ to $2\times10^{-6}$ with slope $1.386293$, against $2\log2=1.386294$, and Richardson extrapolation over $N=100,\dots,800$ at $E=-1.3$ gives the constant $0.140857178$, against $e^{-4g^2}=0.140858421$.

The spectral determinant of each parity Hamiltonian, defined by zeta regularization, equals Braak's function up to a zero-free entire factor~\cite{ReyesBustosWakayama2021,KimotoRBW2021}.

\subsection{Jaynes--Cummings model}
\label{sec:jc}
This model has unbounded couplings and a diagonal kernel. With
\begin{equation}
H=\omega\ad a+\frac{\Delta}{2}\sigma_z+g\bigl(a\sigma_++\ad\sigma_-\bigr),
\end{equation}
the $\sigma_z$ basis gives the bare oscillators $H_{\uparrow\downarrow}=\omega\ad a\pm\Delta/2$ and the couplings $V_{\uparrow\downarrow}=ga$, $V_{\downarrow\uparrow}=g\ad$, relatively bounded with bound zero. The kernel $K=g^2R_\uparrow aR_\downarrow\ad$ is diagonal,
\begin{equation}
\kappa_n(E)=\frac{g^2(n+1)}{\bigl(\omega n+\frac\Delta2-E\bigr)\bigl(\omega(n+1)-\frac\Delta2-E\bigr)}\sim\frac{g^2}{\omega^2n},
\end{equation}
so $\sum_n|\kappa_n|^2<\infty$ and $\sum_n|\kappa_n|=\infty$: $K\in\SS2\setminus\SS1$ and $m=2$. The condition $\kappa_n(E)=1$ reads
\begin{equation}
\Bigl(\omega n+\frac\Delta2-E\Bigr)\Bigl(\omega(n+1)-\frac\Delta2-E\Bigr)=g^2(n+1),
\end{equation}
the secular equation of the $n$th doublet.

\subsection{Rabi--Stark model and the choice of decomposition}
\label{sec:stark}
This model shows that (C3) depends on the decomposition. The Hamiltonian is
\begin{equation}
H=\omega\ad a+\frac{\Delta}{2}\sigma_z+\chi\,\ad a\,\sigma_z+g(a+\ad)\sigma_x,\qquad|\chi|<\omega .
\end{equation}

\emph{$\sigma_x$ basis.} The channels are $H_\pm=\omega\ad a\pm g(a+\ad)$ with coupling $V_{+-}=\Delta/2+\chi\ad a$. With $\bra nR_\mp\ket n=O(1/n)$,
\begin{equation}
\bra n\Sigma(E)\ket n=\frac{\chi^2n}{\omega}+O(1),
\end{equation}
so in this decomposition $\Sigma(E)$ is unbounded and lies outside the scope of (C3). At $\omega=\Delta=1$, $g=0.5$, $\chi=0.3$, $E=-0.7$ the ratio of $\bra n\Sigma\ket n$ to $\chi^2n/\omega$ is $1.004$ at $n=800$, and the singular values of $K(E)$ accumulate at $(\chi/\omega)^2$.

\emph{$\sigma_z$ basis.} The Stark term is diagonal,
\begin{equation}
H_\pm=(\omega\pm\chi)\ad a\pm\frac\Delta2,\qquad V_{+-}=V_{-+}=g(a+\ad),
\end{equation}
with $\Omega_\pm=\omega\pm\chi>0$ for $|\chi|<\omega$. The kernel $K=g^2R_-(a+\ad)R_+(a+\ad)$ lies in $\SS q$ for $q>1$ by Proposition~\ref{prop:upperlin}; its singular values decay as $2g^2/(\Omega_+\Omega_-n)$, so $K\in\SS2\setminus\SS1$ and $m=2$. At the parameters above the two lowest levels from direct diagonalization, $-0.621699628$ and $-0.287186901$, agree with the roots of ${\det}_2[I-K(E)]$ at $N=200$ to $1.2\times10^{-14}$ and $2.8\times10^{-16}$. At $|\chi|=\omega$ one frequency vanishes, at the coupling where a continuum appears in the spectrum~\cite{BraakStark2024,xie2019}.

\subsection{Quadratic channels}
\label{sec:quadratic}
\begin{figure}[t]
\centering
\includegraphics[width=0.6\textwidth]{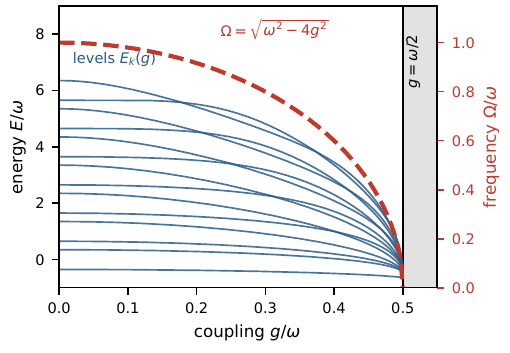}
\caption{Two-photon Rabi model at $\Delta/\omega=0.7$. Thin solid curves: the lowest fourteen levels (left axis). Dashed curve: $\Omega(g)=\sqrt{\omega^2-4g^2}$ (right axis). At $g=\omega/2$ the frequency $\Omega$ vanishes, the channel spectrum becomes a half-line, and the discrete oscillator normal form ends; in the shaded region the channels are unbounded below.}
\label{fig:twophoton}
\end{figure}

This subsection treats channels that reach the normal form through squeezings, and the coupling at which the normal form ends. The two-photon Rabi model is
\begin{equation}
H=\omega\ad a+\frac{\Delta}{2}\sigma_z+g\bigl(a^2+a^{\dagger2}\bigr)\sigma_x .
\end{equation}
The $\sigma_x$ basis gives the scalar coupling $\Delta/2$ and the channels $H_\pm=\omega\ad a\pm g(a^2+a^{\dagger2})$. With $x=(a+\ad)/\sqrt2$ and $p=i(\ad-a)/\sqrt2$,
\begin{equation}
H_\pm=\Bigl(\frac\omega2\pm g\Bigr)x^2+\Bigl(\frac\omega2\mp g\Bigr)p^2-\frac\omega2,
\end{equation}
and a squeezing $S(\xi)$ gives
\begin{equation}
S(\xi)\dg H_\pm S(\xi)=\Omega\,\ad a+\varepsilon_\pm,\qquad\Omega=\sqrt{\omega^2-4g^2}.
\end{equation}
For $g<\omega/2$ the resolvents are Hilbert--Schmidt and $m=1$; the excluded set consists of the two squeezed ladders. At $g=\omega/2$ one coefficient of the Weyl form vanishes and the channel spectrum becomes a half-line. The discrete oscillator normal form and the Schatten estimate of Sec.~\ref{sec:order} end at this point. For $g>\omega/2$ the quadratic form is indefinite and the channel Hamiltonians are unbounded below; choosing a state in one channel gives the same expectation value for the full Hamiltonian. The boundary $g=\omega/2$ is the spectral-collapse point of the two-photon Rabi model~\cite{duan2015,Felicetti2018,Duan2022} (Fig.~\ref{fig:twophoton}); a Bargmann-space spectral determinant was constructed in~\cite{Braak2023TwoPhoton}.

The two-mode model
\begin{equation}
H=\omega(\ad_1a_1+\ad_2a_2)+\frac\Delta2\sigma_z+g(a_1a_2+\ad_1\ad_2)\sigma_x
\end{equation}
becomes, with $b_{1,2}=(a_1\pm a_2)/\sqrt2$, a pair of single-mode two-photon channels with couplings $\pm g/2$, so $\Omega=\sqrt{\omega^2-g^2}>0$ for $|g|<\omega$. Here $N_b=2$, and Propositions~\ref{prop:upper} and~\ref{prop:lowerb} give $m=2$.

\subsection{Three-level emitter}
\label{sec:lambda}
This model has three channels with distinct energies coupled through one mode, so that $\sigma(\Hc)$ differs from the channel spectra. With lower levels $\ket1,\ket2$, an excited level $\ket3$ and energies $\varepsilon_i$,
\begin{equation}
H=\omega\ad a+\sum_{i=1}^3\varepsilon_i\ket i\bra i+(a+\ad)\bigl[g_1\ket1\bra3+g_2\ket2\bra3+{\rm h.c.}\bigr].
\label{eq:lambda}
\end{equation}
The atomic levels are the channels, each a bare oscillator with $\Omega=\omega$. Retaining $\ket1$, the complementary block $\Hc$ is the coupled pair $\{\ket2,\ket3\}$, and
\begin{equation}
K(E)=R_1(E)\,g_1^2(a+\ad)\,\bigl[(\Hc-E)^{-1}\bigr]_{33}\,(a+\ad).
\end{equation}
The vector $v_n=W\dg\ket n=g_1(a+\ad)\ket n$ lies in channel $3$, so $\langle v_n,\Hc v_n\rangle=\langle v_n,H_3v_n\rangle$ and Proposition~\ref{prop:lowerunb} gives $m\ge2$, and Proposition~\ref{prop:upperlin} gives $m\le2$, so $m=2$.

\begin{figure}[t]
\centering
\includegraphics[width=0.65\textwidth]{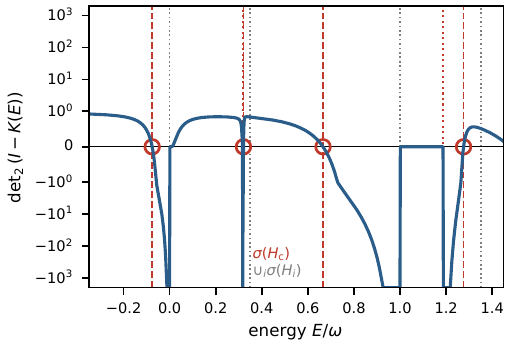}
\caption{Three-level emitter at $\omega=1$, $(\varepsilon_1,\varepsilon_2,\varepsilon_3)=(0,0.35,2.1)$, $g_1=0.45$, $g_2=0.30$, truncation $N=45$. Solid curve: ${\det}_2[I-K(E)]$ on an $\operatorname{arsinh}$ scale; circles: eigenvalues of $H$; grey dotted lines: $\sigma(H_1)$; red dotted lines: $\sigma(\Hc)$. The curve is analytic across $\sigma(H_2)=\{0.35,1.35\}$.}
\label{fig:lambda}
\end{figure}

At $\omega=1$, $(\varepsilon_1,\varepsilon_2,\varepsilon_3)=(0,0.35,2.1)$, $g_1=0.45$, $g_2=0.30$, the excluded set \eqref{eq:domain} consists of
\begin{equation}
\sigma(H_1)=\{0,1,2,\dots\},\qquad
\sigma(\Hc)=\{0.31657171,\ 1.18678948,\ 2.08627055,\dots\}.
\end{equation}
The points of $\sigma(H_2)$ remain regular for the retained-channel kernel, while $\sigma(H_1)\cup\sigma(\Hc)$ forms the singular set. The zeros
\begin{equation}
-0.0758184286,\quad 0.3204001586,\quad 0.6657783536,\quad 1.2756877552
\end{equation}
agree with direct diagonalization to $10^{-15}$ (Fig.~\ref{fig:lambda}). For $\varepsilon_1=\varepsilon_2$ the combination $g_2\ket1-g_1\ket2$ decouples and the problem reduces to two channels~\cite{Wu1996}.

\subsection{Dicke model}
\label{sec:dicke}
This model has a number of channels that grows with the number of qubits and unbounded couplings along a chain. With $N$ qubits, $j=N/2$ and collective spin operators $J_z,J_\pm$,
\begin{equation}
H=\omega\ad a+\omega_0J_z+\frac{g}{\sqrt N}(a+\ad)(J_++J_-)
\end{equation}
on the symmetric subspace. The $J_z$ eigenbasis gives $r=N+1$ channels $H_\mu=\omega\ad a+\omega_0\mu$, $\mu=-j,\dots,j$, coupled along a chain,
\begin{equation}
V_{\mu,\mu\pm1}=\frac{g}{\sqrt N}\sqrt{j(j+1)-\mu(\mu\pm1)}\,(a+\ad).
\end{equation}
Retaining $\mu=-j$, the vector $W\dg\ket n$ lies in channel $\mu=-j+1$, Proposition~\ref{prop:lowerunb} gives $m\ge2$, and Proposition~\ref{prop:upperlin} gives $m\le2$, so $m=2$. The excluded set is $\sigma(H_{-j})\cup\sigma(\Hc)$; at $N=3$, $\omega=1$, $\omega_0=0.8$, $g=0.5$ the three lowest points of $\sigma(\Hc)$ are $-0.6412430358$, $-0.3584133914$, $0.3218448403$. The roots of ${\det}_2[I-K(E)]$ at $N_{\rm F}=45$ agree with direct diagonalization to $10^{-14}$ for $N=3$ and $N=5$, where $(\Hc-E)^{-1}$ is evaluated as the resolvent of the coupled block. Explicit solutions of the Dicke model are given in~\cite{BraakDicke2013,HeDuanChen2015}.

\subsection{Arrays of inhomogeneous qubits}
\label{sec:qubits}
This subsection evaluates the criterion for $N$ qubits with individual couplings, using Proposition~\ref{prop:BS} and the parity symmetry. The Hamiltonian is
\begin{equation}
H=\omega\ad a+\sum_{i=1}^N\Bigl[\frac{\Delta_i}{2}\sigma_{i,z}+g_i(a+\ad)\sigma_{i,x}\Bigr].
\label{eq:Nqubit}
\end{equation}
In the joint eigenbasis $\ket s$, $s\in\{\pm1\}^N$, of the $\sigma_{i,x}$, each channel is a displaced oscillator,
\begin{equation}
H_s=\omega\ad a+G_s(a+\ad),\qquad G_s=\sum_ig_is_i ,
\end{equation}
with ladder
\begin{equation}
E^{(0)}_{s,n}=n\omega-\frac{G_s^2}{\omega},
\label{eq:ladder}
\end{equation}
and $\sigma_{i,z}$ couples $s$ to $f_i(s)$, the configuration with $s_i$ reversed, with amplitude $\Delta_i/2$. In the polaron frames $D(-G_s/\omega)$ the coupling block from $t$ to $s$ is
\begin{equation}
\frac{\Delta_i}{2}\,D\Bigl(\frac{G_s-G_t}{\omega}\Bigr),\qquad t=f_i(s),\qquad G_s-G_t=2g_is_i .
\end{equation}

\emph{Parity reduction.} The operator $\mathcal P=\prod_i\sigma_{i,z}\,(-1)^{\ad a}$ commutes with $H$ and maps $s\to-s$. Choosing the representatives $S_+=\{s:s_1=+1\}$, the parity-$p$ subspace is described on $\bigoplus_{s\in S_+}\mathcal H_s$ by the diagonal part $H_0$ with entries \eqref{eq:ladder} and the coupling $V_p$ with blocks
\begin{equation}
(V_p)_{s,t}=\frac{\Delta_i}{2}\,D\Bigl(\frac{2g_is_i}{\omega}\Bigr)\times
\begin{cases}1,& f_i(s)=t\in S_+,\\ p\,\Pi,& f_i(s)=-t,\ t\in S_+ .\end{cases}
\end{equation}
For $N\ge2$ we have $f_i(s)\neq\pm s$, so the diagonal blocks of $V_p$ vanish and
\begin{equation}
\tr\bigl[P_NR_0(E)V_pP_N\bigr]=0
\label{eq:traceless}
\end{equation}
for truncations $P_N$ compatible with the channels. With $A_p(E)=R_0(E)V_p\in\SS2$, Proposition~\ref{prop:BS} gives
\begin{equation}
E\in\sigma(H|_{\mathcal P=p})\iff\mathcal D_p(E)=0,\qquad
\mathcal D_p(E)={\det}_2\bigl[I+A_p(E)\bigr]=\lim_{N\to\infty}\det\bigl[I+P_NA_p(E)P_N\bigr],
\end{equation}
and Lemma~\ref{lem:mero} with $t=0$ shows that $\mathcal D_p$ is meromorphic with poles on the ladders \eqref{eq:ladder} only. For $N=1$ the single flip maps $s$ to $-s$, the block $\frac\Delta2D(2\lambda)\,p\,\Pi$ sits on the diagonal, $A_p=pX$ with $X$ of \eqref{eq:X}, and
\begin{equation}
\mathcal D_p(E)\,e^{p\tau(E)}=D_{-p}(E),
\end{equation}
with $D_\pm$ of \eqref{eq:Dpm}. In the polaron basis the truncation obeys
\begin{equation}
\det\bigl[I+P_NA_p(E)P_N\bigr]=\frac{\det\bigl[P_N(H_0+V_p-E)P_N\bigr]}{\det\bigl[P_N(H_0-E)P_N\bigr]} .
\end{equation}

\begin{figure}[t]
\centering
\includegraphics[width=0.65\textwidth]{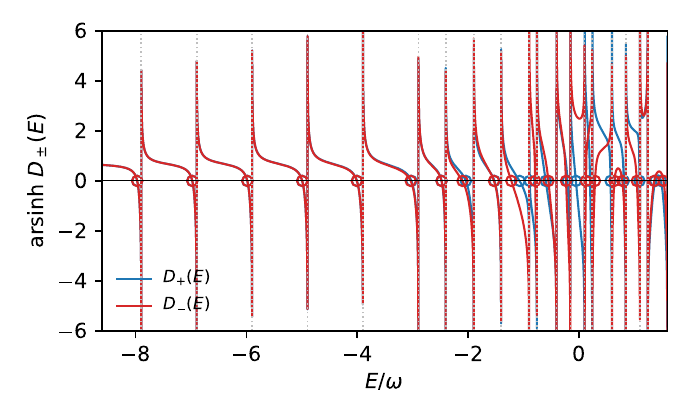}
\caption{Parity determinants $\mathcal D_\pm(E)$ for three qubits at $\omega=1$, $g=(0.63,0.97,1.21)$, $\Delta=(0.7,1.0,1.3)$, on an $\operatorname{arsinh}$ scale. Circles: eigenvalues of each parity block from direct diagonalization; dotted lines: ladders \eqref{eq:ladder}.}
\label{fig:nqubit}
\end{figure}

For $N=3$, $\omega=1$, $g=(0.30,0.45,0.60)$, $\Delta=(0.7,1.0,1.3)$, the zeros of $\mathcal D_\pm$ below $E=4$ number $21$ in each parity and agree with direct diagonalization of each parity block to $2\times10^{-14}$ at $N=60$ and $N=90$. Figure~\ref{fig:nqubit} shows $\mathcal D_\pm$ at stronger coupling. Explicit spectral functions are known for one and two arbitrary qubits~\cite{Braak2011,TwoQubitRabi,Duan2015TwoQubit} and for the Dicke model~\cite{BraakDicke2013,HeDuanChen2015}.

\section{Conclusion}
\label{sec:conclusion}

A finite channel decomposition, with one channel retained and the remaining channels treated as a single coupled block, turns the spectral problem into a determinant condition on the retained channel. Under conditions stated on the channel operators and their couplings, namely self-adjoint channels bounded below, couplings relatively bounded with relative bound below one, a bounded self-energy, and a reduced kernel in a Schatten class, the eigenvalues outside the spectra of the retained channel and the coupled block are exactly the zeros of a regularized Fredholm determinant, with the eigenspace dimension preserved. The strict negativity of the energy derivative of the Schur complement makes the order of each zero equal to the multiplicity of the eigenvalue. For oscillator channels the normal form makes the channel resolvents diagonal, and the reduced kernel carries two resolvent factors, which halves its Schatten index relative to the full-space Birman--Schwinger kernel. The regularization order is the smallest integer above half the number of modes for bounded couplings and equals two for linear couplings to a single mode, and matching lower bounds show that both orders are optimal for bounded displacement and squeezing couplings and for linear single-mode couplings.

The models separate these ingredients. The Friedrichs--Lee model, with a continuum channel and a kernel of finite rank, lies outside the oscillator class and inside the criterion. The quantum Rabi and two-photon models give trace-class kernels; in the Rabi model the determinant factorizes into parity sectors, and a residue condition on the excluded ladder reproduces the Judd solutions. Unbounded couplings in the Jaynes--Cummings, Rabi--Stark, three-level and Dicke models, and a second oscillator mode in the two-mode model, give Hilbert--Schmidt kernels. In the Rabi--Stark model the conditions select the decomposition into bare qubit states, and in the two-photon model the oscillator normal form ends at the spectral-collapse coupling. The three-level emitter makes the spectrum of the coupled complementary block visible, and the Dicke model and arrays of inhomogeneous qubits extend the construction to larger channel graphs.

\end{document}